\documentclass[preprint,12pt,3p]{elsarticle}
\usepackage[utf8]{inputenc}
\usepackage{placeins}
\usepackage{color}
\usepackage{hyperref}

\journal{Computers and Security}

\begin{document}
\begin{frontmatter}

\title{Early-Stage Malware Prediction Using Recurrent Neural Networks}

\author[label1]{Matilda Rhode\corref{cor1}}
\address[label1]{School of Computer Science and Informatics, Cardiff University}
\cortext[cor1]{Corresponding author}
\ead{rhodem@cardiff.ac.uk}

\author[label1]{Pete Burnap}
\ead{burnapp@cardiff.ac.uk}

\author[label2]{Kevin Jones}
\address[label2]{Airbus Group}
\ead{kevin.jones@airbus.com}

\begin{abstract}
     Static malware analysis is well-suited to endpoint anti-virus systems as it can be conducted quickly by examining the features of an executable piece of code and matching it to previously observed malicious code. However, static code analysis can be vulnerable to code obfuscation techniques. Behavioural data collected during file execution is more difficult to obfuscate, but takes a relatively long time to capture - typically up to 5 minutes, meaning the malicious payload has likely already been delivered by the time it is detected.
     
     In this paper we investigate the possibility of predicting whether or not an executable is malicious based on a short snapshot of behavioural data. We find that an ensemble of recurrent neural networks are able to predict whether an executable is malicious or benign within the first 5 seconds of execution with 94\% accuracy. This is the first time general types of malicious file have been predicted to be malicious during execution rather than using a complete activity log file post-execution, and enables cyber security endpoint protection to be advanced to use behavioural data for blocking malicious payloads rather than detecting them post-execution and having to repair the damage.
\end{abstract}

\begin{keyword}
malware detection \sep intrusion detection \sep recurrent neural networks \sep machine learning  \sep deep learning
\end{keyword}

\end{frontmatter}

\section{Introduction}

Automatic malware detection is necessary to process the rapidly rising rate and volume of new malware being generated. Virus Total, a free tool which can be used to evaluate whether files are malicious, regularly approaches one million new, distinct files for analysis each day\footnote{0.935 million on 2nd December 2017}\cite{virusTotalStats}. 

Commonly, automatic malware detection used in anti-virus systems compares (features extracted from) the code of an incoming file to a known list of malware signatures. However, this form of filtering using static data is unsuited to detecting completely new (``zero-day") malware unless it shares code with previously known strains \cite{vinod2009survey}. Obfuscating the code, now common practice among malware authors, can even enable previously seen malware to escape detection \cite{obfuscationTechniques}.

Malware detection research has evolved to respond to the inadequacies of static detection. Behavioural analysis (dynamic analysis) examines a sample file in a virtual environment whilst it is being executed. Behavioural analysis approaches assume that malware cannot avoid leaving a measurable footprint as a result of the actions necessary for it to achieve its aims. However, executing the malware incurs a time penalty by comparison with static analysis. Whilst dynamic data can lead to more accurate and resilient detection models than static data (\cite{Nataraj}, \cite{Damodaran2017}, \cite{grosse2016adversarial}), in practice behavioural data is rarely used in commercial endpoint anti-virus systems due to this time penalty. It is inconvenient and inefficient to wait for several minutes whilst a single file is analysed, and ultimately, the malicious payload has likely been delivered by the end of the analysis window so the opportunity to block malicious actions has been missed. 

To avoid waiting, some approaches monitor ``live" activity on the local network or the machine. These detection systems tend either to look for traits that signify a particular type of malware (e.g. ransomware) or to flag deviations from a baseline of ``normal" behaviour. These two approaches suffer from specific flaws. Searching for particular behaviours is analogous to the traditional methods of comparing incoming files with known variants, and may miss detecting new types of malware. Whilst anomaly detection is prone to a high false-positive rate as any activity that deviates from a ``normal" baseline is deemed malicious. In practice anomalous activity is often investigated by human analysts, making the model vulnerable to exploitation. An attacker could bring about lots of anomalous behaviour such that the human analysts are flooded with investigation requests, reducing the chances of the activity created by the attack itself from being detected.

We propose a behaviour-based model to predict whether or not a file is malicious using the first few seconds of file execution with a view to developing a tool that could be incorporated into an end-point solution. Though general malicious and benign files comprise a wide range of software and potential behaviours, our intuition is that malicious activity begins rapidly once a malicious file begins execution because this reduces the overall runtime of the file and thus the window of opportunity for being disrupted (by a detection system, analyst, or technical failure). As far as we are aware this is the first paper attempting to predict malicious behaviour for various types of malware based on early stage activity. 

We feed a concise feature set of file machine activity into an ensemble of recurrent neural networks and find that we achieve a 94\% accurate detection of benign and malicious files 5 seconds into execution. Previous dynamic analysis research collects data for around 5 minutes per sample.

The main contributions of this paper are:
\begin{enumerate}
    \item We propose a recurrent neural network (RNN) model to predict malicious behaviour using machine activity data and demonstrate its capabilities are superior to other machine learning solutions that have previously been used for malware detection
    \item We conduct a random search of hyperparameter configurations and provide details of the configurations leading to high classification accuracy, giving insight into the methods required for optimising our malware detection model
    \item We investigate the capacity of our model to detect malware families and variants which it has not seen previously - simulating `zero day' and advanced persistent threat (APT) attacks that are notoriously difficult to detect
    \item We conduct a case-study using 3,000 ransomware samples and show that our model has high detection accuracy (94\%) at 1 second into execution without prior exposure to examples of ransomware, and investigate the combinations of features most relevant to the model decisions
\end{enumerate}

\section{Related Work}

Automatic malware detection models typically use either code or behaviour based features to represent malicious and benign samples. Each of these approaches has its benefits and drawbacks, such that research continues to explore detection methods using both kinds of data. 

Hybrid approaches use both static and dynamic data, closer approximating the methods used by anti-virus engines; why analyse the behaviour of a file if it matches a known malware signature? But unless static detection is used purely to filter out known malwares, any dependence on static methods in a hybrid approach leaves the model open to the same weaknesses as a purely static model. 

\paragraph{Static data} Static data, derived directly from code, can be collected quickly. Though signature-based methods fail to detect obfuscated or entirely new malware, researchers have extracted other features for static detection. Saxe and Berlin \cite{saxeBerlin} distinguish malware from benignware using a deep feed-forward neural network with a true-positive rate of 95.2\% using features derived from code. However, the true-positive rate falls to 67.7\% when the model is trained using files only seen before a given date and tested using those discovered for the first time after that date, indicating the weakness of static methods in detecting completely new malwares. Damodoran et al. \cite{Damodaran2017} conducted a comparative study of static, behavioural and hybrid detection models for malware detection and found behavioural data to give the highest area under the curve (AUC) value, 0.98, using Hidden Markov Models with a dataset of 785 samples. Additionally, Grosse et al.\cite{grosse2016adversarial} show that, in the case of Android software, static data can be obfuscated to cause a classifier previously achieving 97\% accuracy to fall as low as 20\% when classifying obfuscated samples. Training using obfuscated samples allowed a partial recovery of accuracy, but accuracy did not improve beyond random chance. 

\paragraph{Dynamic data} Methods using dynamic data assume that malware must enact the behaviours necessary to achieve their aims. Typically, these approaches capture behaviours such as API calls to the operating system kernel. Tobiyama et al.\cite{malwareProcessBehaviour}  use RNNs to extract features from 5 minutes of API call log sequences which are then fed into a convolutional neural network to obtain 0.96 AUC score with a dataset of 170 samples. Firadusi et al. \cite{FirdausiBehaviour} compare machine learning algorithms trained on API calls and achieve an accuracy of 96.8\% using correlation-based feature selection and a J48 decision tree. The 250 benign samples used for the experiment are all collected from the WindowsXP System32 directory, which is likely to give a higher degree of homogeneity than benign software encountered in the wild. Ahmed et al. \cite{ahmed2009using} detect malware using API call streams and associated metadata with a Naive Bayes classifier, achieving 0.988 AUC, again with the 100 benign samples being WindowsXP 32-bit system files. Both Tian et al. \cite{tian2010differentiating} and Hansen et al. use Random Forests trained on API calls and associated metadata to achieve 97\% accuracy and a 98\% F-Score respectively. Huang and Stokes \cite{Huang:2016} achieve the highest accuracy in the literature, 99.64\%, using System API calls and features derived from those API calls using a shallow feed-forward neural network. Table~\ref{dynamic_times} outlines the dataset sizes and recording time for the related literature. The median dataset size for binary classification is 1,300 samples. Huang and Stokes \cite{Huang:2016} and Pascanu et al. \cite{PascanuMalwareRNN} are outliers with much larger datasets, both obtained through access to the corpus of samples held privately by the authors' companies. The majority of research does not mention a time-cap on file execution, in these cases we may presume that the files are executed until activity stops. The median data capture time frame for those reported is 5 minutes (see Table~\ref{dynamic_times}). 

\begin{table}[hb]
    \small
    \centering
    \begin{tabular}{l|p{1.6cm}|p{1.6cm}|p{6cm}}
        \hline
        Ref. & Malicious samples & Benign samples & Reported time collecting dynamic data \\\hline\hline
        \multicolumn{4}{c}{Binary classification}\\\hline\hline
        \cite{malwareProcessBehaviour} & 81 & 69 & 5 minutes \\\hline
        \cite{FirdausiBehaviour} & 220 & 250 & No time cap mentioned - implicit full execution \\\hline
        \cite{ahmed2009using} & 416 & 100 & No time cap mentioned - implicit full execution \\\hline
        \cite{Damodaran2017} & 745 & 40 & Fixed time and 5-10 minutes mentioned but overall time cap not explicitly stated\\\hline
        \cite{tian2010differentiating} & 1,368 & 465 & 30 seconds \\\hline
        \cite{HansenForest}\footnote{Dataset sizes reported for binary classification problem, not malware family detection} & 5,000 & 837 & 3.33 minutes (200 seconds)\\\hline
        \cite{PascanuMalwareRNN} & 25,000 & 25,000 & At least 15 steps - exact time unreported \\\hline
        \cite{Huang:2016} &2.85m & 3.65m & No time cap mentioned - implicit full execution \\\hline\hline
        \multicolumn{4}{c}{Malware family classification}
        \\\hline\hline
        \cite{kolosnjaji2016deep} & 4,753 & n/a &  No time cap mentioned - implicit full execution \\\hline
    \end{tabular}
    \caption{Reported data sample sizes and times collecting dynamic behavioural data per sample}
    \label{dynamic_times}
\end{table}

\paragraph{Time-efficiency dynamic analysis methods} Existing methods to reduce dynamic data recording time focus on efficiency. The core concept is only to record dynamic data if it will improve accuracy, either by omitting some files from dynamic data collection or by stopping data collection early. Shibahara et al. \cite{ShibaharaEfficientDynamics} decide when to stop analysis for each sample based on changes in network communication, reducing the total time taken by 67\% compared with a ``conventional" method that analyses samples for 15 minutes each. Neugschwandtner et al.~\cite{neugschwandtner2011forecast} used static data to determine dissimilarity to known malware variants using a clustering algorithm. If the sample is sufficiently unlike any seen before, dynamic analysis is carried out. This approach demonstrated an improvement in classification accuracy by comparison with randomly selecting which files to dynamically analyse, or selecting based on sample diversity. Similarly, Bayer et al. \cite{bayer2010improving} create behavioural profiles to try and identify polymorphic variants of known malware, reducing the number of files undergoing full dynamic analysis by 25\%. Approaches to date still allow some files to be run for a long dynamic execution time, whereas here we investigate a blanket cut-off of dynamic analysis for all samples, with a view to this analysis being run in an endpoint anti-virus engine. 

\paragraph{RNNs for malware detection} We propose using a recurrent neural network (RNN) for predicting malicious activity as as they are able to process time-series data, thus capturing information about change over time as well as the raw input feature values. Kolsnaji et al \cite{Kolosnjaji2016} sought to detect malware families with deep neural networks, including recurrent networks, to classify malware into families using API call sequences. By combining a convolutional neural network with long-short-term memory (LSTM) cells, the authors were able to attain a recall of 89.4\%, but do not address the binary classification problem of distinguishing malware from benignware. Pascanu et al. \cite{PascanuMalwareRNN} did conduct experiments into whether files were malicious or benign using RNNs and Echo State Networks. The authors found that Echo State Networks performed better with an accuracy of around 95\% (error rate of 5\%) but did not attempt to predict malicious behaviour from initial execution. 

\paragraph{Ransomware detection} In Section~\ref{ransomware} we test our model on a corpus of 3,000 ransomware samples. Early prediction is particularly useful for types of malware from which recovery is difficult and/or costly. Ransomware encrypts user files and withholds the decryption key until a ransom is paid to the attackers. This type of attack cannot be remedied without financial loss unless a backup of the data exists. Recent work on ransomware detection by Scaife et al. \cite{scaife2016cryptolock} uses features from file system data, such as whether the contents appears to have been encrypted, and number of changes made to the file type. The authors were able to detect and block all of the 492 ransomware samples tested with less than 33\% of user data being lost in each instance. Continella et al. \cite{continella2016shieldfs} propose a self-healing system, which detects malware using file system machine activity (such as read/write file counts), the authors were able to detect all 305 ransomware samples tested, with a very low false-positive rate. These two approaches use features selected specifically for their ability to detect ransomware, but this requires knowledge of how the malware operates. Our approach seeks to use features which can be used to detect any malware family, including those which have not been seen before. That is to say, we will demonstrate the effectiveness of detecting ransomware without dependence on ransomware-specific training data. The key purpose of this final experiment is to show that our general model of malware detection is able to detect general types of malware as well as time-critical samples such as ransomware.

\section{Methods}

Dynamically collected data is more robust to obfuscation methods than statically collected data (\cite{Damodaran2017}, \cite{grosse2016adversarial}), but dynamic collection takes longer. In order to advance malware detection to a more predictive model that can respond in seconds we propose a model which uses only short sequences of the initial dynamic data to investigate whether this is sufficient to judge a file as malicious with a high degree of accuracy.

We use 10 machine activity data metrics as feature inputs to the model. We take a snapshot of the metrics every second for 20 seconds whilst the sample executes, starting at 0s, such that at 1s, we have two feature sets or a sequence length of 2. Though API calls to the operating system kernel are the most popular behavioural features used in dynamic malware detection, there are several reasons why we have chosen machine activity features as inputs to the model instead. Firstly, recent work has shown that API calls are vulnerable to manipulation, causing neural networks to misclassify samples (\cite{CPE4023}, \cite{RosenbergSRE17}). As Burnap et al. \cite{somburnap} argue``malware cannot avoid leaving a behavioural footprint" of machine activity, future work will necessarily examine the robustness of machine activity to adversarial crafting, but this is outside the scope of this paper. A key advantage of continuous data such as machine activity metrics is that the model is able to infer information from completely unseen input values; any unseen data values in the test set will still have numerical relevance to the data from the training set as it will have a relative value that can be mapped onto the learned model. API calls on the other hand are categorical, such the meaning of unseen API call cannot be interpolated against existing data. Practically, categorical features require an input vector with a placeholder for each category to record whether it is present or not. Hundreds or even thousands (\cite{Huang:2016}) of API calls can be collected, leading to a very large input vector, which in turn makes the model slower to train. Being categorical, any API calls not present in the training data will have no placeholder in the input vector at the classification stage even if they appear in later test samples. The machine activity data we collected are continuous numeric values, allowing for a large number of different machine states to be represented in a small vector of size 10.

\begin{figure}[ht]
\centering
\includegraphics[width=0.75\textwidth]{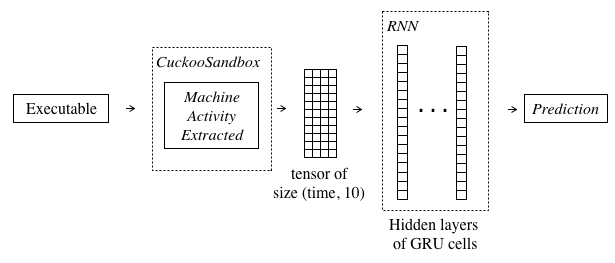}
\caption{High-level model overview}
\label{system}
\end{figure}

As illustrated in Figure \ref{system}, to collect our activity data we executed Portable Executable (PE) samples using Cuckoo Sandbox \cite{cuckooSandbox}, a virtualised sandboxing tool. While executing each sample we extracted machine activity metrics using a custom auxiliary module reliant on the Python Psutil library\cite{psutil}. The metrics captured were: system CPU usage, user CPU use, packets sent, packets received, bytes sent, bytes received, memory use, swap use, the total number of processes currently running and the maximum process ID assigned.

\FloatBarrier
As the data are sequential, we chose an algorithm capable of analysing sequential data. Making use of the time-series data means that the rate and direction of change in features as well as the raw values themselves are all inputs to the model. Recurrent Neural Networks (RNNs) and Hidden Markov Models are both able to capture sequential changes, but RNNs hold the advantage in situations with a large possible universe of states and memory over an extended chain of events \cite{lipton15critical}, and are therefore better suited to detecting malware using machine activity data. 

RNNs can create temporal depth in the same way that neural networks are deep when multiple hidden layers are used. Until the development of the LSTM cell by Hochreiter and Schmidhuber in 1997, RNNs performed poorly in classifying long sequences, as the updates required to tune the weights between neurons would tend to vanish or explode \cite{bengio1994learning}. LSTM cells can hold information back from the network until such a time as it is relevant or ``forget" information, thus mitigating the problems surrounding weight updates. The success of LSTM has prompted a number of variants, though few of these have significantly improved on the classification abilities of the original model \cite{greff2016lstm}. Gated Recurrent Units (GRUs) \cite{cho2014GRU}, however, have been shown to have comparable classification to LSTM cells, and in some instances can be faster to train \cite{Chung2014GRULSTM}, for this potential training speed advantage, we use GRU units. 

An appropriate architecture and learning procedure of a neural network is usually integral to a successful model. These attributes are captured by hyperparameter settings, which are often hand-crafted. Due to the rapid evolution of malware, we anticipate that the RNN should be re-trained regularly with newly discovered samples, thus the architecture may need to change too. As it needs to be carried out multiple times, this process should be automated. We chose to conduct a random search of the hyperparameter space as it can easily be parallelised (unlike a grid search), it is trivial to implement, and has been found to be more efficient at finding good configurations than grid search \cite{bergstra2012random}. We chose the configuration which performed best on a 10-fold cross-validation over the training set for our final model configuration, the hyperparameter search space and final configuration is detailed in Table~\ref{hyperparam_table} for reproducibiltiy.

\begin{table}[h]
    \centering
    \small
    \begin{tabular}{l|l|l}
        \hline
        Hyperparameter & Possible values & Best configuration\\\hline\hline
        Depth & 1, 2, 3 & 3 \\\hline
        Bidirectional & True, False & True \\\hline
        Hidden neurons & 1 -- 500 & 74 \\\hline
        Epochs & 1 -- 500 & 53 \\\hline
        Dropout rate & 0 -- 0.5 (0.1 increments) & 0.3 \\\hline
        Weight regularisation & None, \(l1\), \(l2\), \(l1\) and \(l2\) & \(l2\) \\\hline
        Bias regularisation & None, \(l1\), \(l2\), \(l1\) and \(l2\) & None \\\hline
        Batch size & 32, 64, 128, 256 & 64 \\\hline
    \end{tabular}
    \caption{Possible hyperparameter values and the hyperparameters of the best-perfoming configuration on the training set}
    \label{hyperparam_table}
\end{table}

\section{Dataset}

\subsection{Samples}

We initially obtained 1,000 malicious and 600 ``trusted" Windows7 executables from VirusTotal \cite{virusTotal} along with 800 trusted samples from the system files of a fresh Windows7 64-bit installation. We then downloaded a further 4,000 Windows 7 applications from popular free software sources, such as Softonic \cite{softonic}, PortableApps \cite{portableapps} and SourceForge \cite{sourceforge}. We included the online download files as they are a better representation the typical workload of an anti-virus system than Windows system files.

We used the VirusTotal API \cite{virusTotal} as a proxy to label the downloaded software as benign or malicious. VirusTotal runs files through around 60 anti-virus engines and reports the number of engines that detected the file as malicious. Similar to \cite{saxeBerlin}, for malicious samples, we omitted any files that were deemed malicious by less than 5 engines in the VirusTotal API as the labelling of these files is contentious. Files not labelled as malicious by any of the anti-virus engines were deemed 'trusted' as there is no evidence to suggest they are malware. We therefore consider these as benign samples. This has the limitation of not detecting previously unseen malware but our samples are selected from an extended time period historically so it is likely that it would be reported as malware at some point in this period if it were actually malicious.

The final dataset comprised 2,345 benign and 2,286 malicious samples, which is consistent with dataset sizes in this field of research  e.g. (\cite{FirdausiBehaviour}, \cite{tian2010differentiating}, \cite{yuan2016droiddetector}, \cite{ahmed2009using}, \cite{Damodaran2017}, \cite{malwareProcessBehaviour}, \cite{ImranHMM}). We used a further 2,876 ransomware samples obtained from the VirusShare online malware repository \cite{VirusShare} for the ransomware case study in Section~\ref{ransomware}. 

We were also able to extract the date that VirusTotal had first seen each file and the families and variants that each anti-virus engine classified the malware samples. The dates that the files were first seen ranged from 2006 to 2017. We split the test and training set files according to the date first seen to mimic the arrival of completely new software. The training set only comprised samples first seen by VirusTotal before 11:15 on 10th October 2017 and the test set only samples after this date, which produced a test set of 500 samples (206 trusted and 316 malicious). We choose this date and time as it gave a  number of each malicious and benign samples that is is line with the sample size in the existing literature. 

The total instances of the different malware families is documented in Table~\ref{malwareDemographics}. The ``disputed" class represents those malware for which a family could not be determined because the anti-virus engines did not produce a majority vote in favour of one type. We also found the precise variants where possible, and have listed the numbers of advanced persistent threat malware (APTs) and ransomware in each category as APTs are notoriously difficult for static engines to detect and the ransomware case-study in Section~\ref{ransomware} required removal of all ransomware from the training set.

\begin{table}[h]
    \small
    \centering
    \begin{tabular}{l|r}
    \hline
        Family      & Total (apt)(ransomware) \\\hline\hline
        Trojan             &  1,382 (0)(76)\\
        Virus              &   407 (20)(56)\\
        Adware             &   180 (0)(51)\\
        Backdoor           &   123 (7)(0)\\
        Bot                &   76\\
        Worm               &    24\\
        Rootkit            &    11\\
        Disputed           &   83\\\hline
        \textbf{Total}  & \textbf{2,239}\\\hline
    \end{tabular}
    \caption{Number of instances of different malware families in dataset}
    \label{malwareDemographics}
\end{table}

\subsection{Input Features}

\begin{figure}[h]
    \centering
    \includegraphics[width=0.75\textwidth]{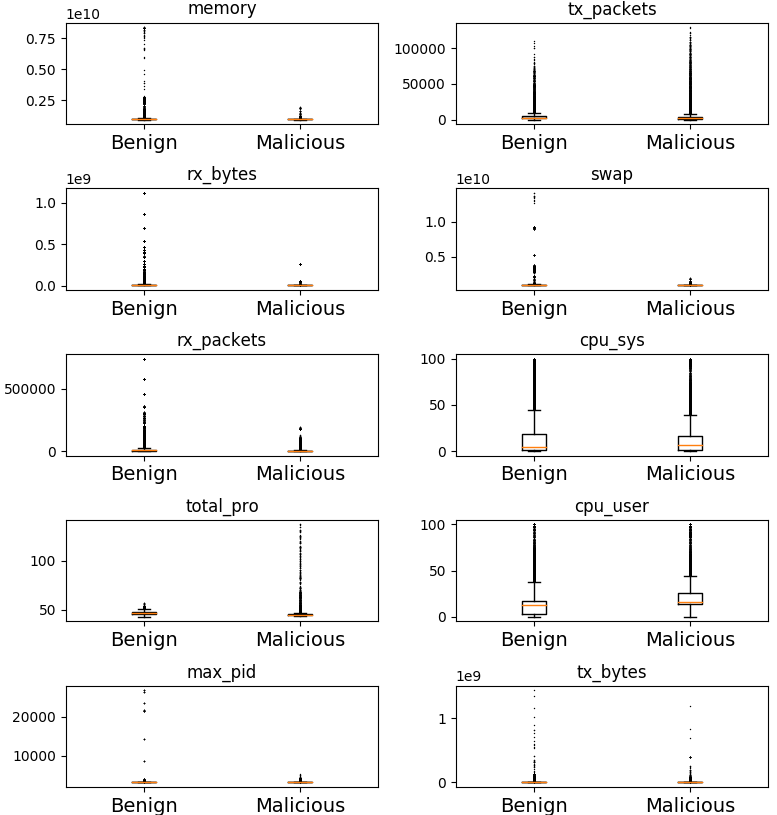}
    \caption{Frequency distributions of input features for benign and malicious samples}
    \label{data_descriptors}
\end{figure}

Table~\ref{descriptor_table} outlines the minimum and maximum values of the 10 inputs we collected for malware and benignware respectively. Though the inter-quartile ranges of values are generally similar (See Figure~\ref{data_descriptors}) The benign data sees a far greater number of outliers in RAM use (memory and swap) and packets being received. The malicious data has a large number of outliers in total number of processes, but the benign samples have outliers in the maximum assigned process ID, indicating that malicious files in this dataset try to carry out lots of longer processes simultaneously, whereas benign files will carry out a number of quick actions in succession. 

\begin{table}[h]
    \small
    \centering
    \begin{tabular}{l||l|l||l|l}
    \hline
        &\multicolumn{2}{c||}{Benign} & \multicolumn{2}{c}{Malicious} \\\hline
        & Min. & Max. & Min. & Max. \\\hline\hline
        Total Processes & 43 & 57 & 44 & 137 \\\hline
        Max. Process ID & 3,020 & 26,924 & 3,020 & 5,084  \\\hline
        CPU User (\%)& 0 & 100 & 0 & 100  \\\hline
        CPU System (\%) & 0 & 100 & 0 & 100  \\\hline
        Memory Use (MB) & 941 & 8,387 & 939 & 1,957 \\\hline
        Swap Use (MB) & 941 & 14,040 & 941 & 1,956 \\\hline
        Packets Sent (000s) & 0.3 & 110 & 0.3 & 129  \\\hline
        Packets Received (000s) & 2.9 & 737 & 2.9 & 192 \\\hline
        Bytes Received (MB)& 4 & 1,116 & 4 & 266 \\\hline
        Bytes Sent (MB)& 0.4 & 1,434 & 0.4 & 1,188 \\\hline
    \end{tabular}
    \caption{Minimum and maximum values of each input feature for benign and malicious samples}
    \label{descriptor_table}
\end{table}

\paragraph{Data preprocessing}
Prior to training and classification, we normalise the data to improve model convergence speed in training. By keeping data between 1 and -1, the model is able to converge more quickly, as the neurons within the network operate within this numeric range \cite{lecun2012efficient}. We achieve this by normalising around the zero mean and unit variance of the training data. For each feature, \(i\) , we establish the mean, \(\mu_{i}\), and variance, \(\sigma_{i}\), of the training data. These values are stored, after which every feature, \(x_{i}\) is scaled: \[\frac{x_{i} - \mu_{i}}{\sigma_{i}}\]

\FloatBarrier

\section{Experimental Results}

For reproducibility, the code used to implement the following experiments can be found at \url{https://github.com/mprhode/malware-prediction-rnn}. Information on the data supporting the results presented here, including how to access them, can be found in the Cardiff University data catalogue at \url{http://doi.org/10.17035/d.2018.0050524986}. We used Keras \cite{chollet2015keras} to implment the RNN experiments, ScikitLearn \cite{scikit-learn} to implement all other machine lerning algorithms and trained the models using an Nvidia GTX1080 GPU. The Virtual Machine used 8GB RAM, 25GB storage, and a single CPU core running 64-bit Windows 7. We installed Python 2.7 on the machine along with a free office software suite (LibreOffice), browser (Google Chrome) and PDF reader (Adobe Acrobat). The virutal machine was restarted between each sample execution to ensure that malicious and benign files alike began from the same machine set-up.

\subsection{Hyperparameter configuration}
Each layer of a neural network learns an abstracted representation of the data fed in from the previous layer. There must be a sufficient number of neurons in each layer and a sufficient number of layers to represent the distinctions between the output classes. The network can also learn to represent the training data too closely, causing the model to overfit. Choosing hyperparameters is about finding a nuanced, but generalisable representation of the data. Table~\ref{hyperparam_table} details the search space and final hyperparameters selected for the models in the later experiments. Although there are only 8 parameters to tune, but there are 576 million different possible configurations. As well as the hyperparameters above, we randomly select the time into execution of data. Although the goal is to find the best classifier for the shortest amount of time, selecting an arbitrary time such as 5 or 10 seconds into file execution may only produce models capable of high accuracy at that sequence length. We do not know whether a model will increase monotonically in accuracy with more data or peak at a particular time into the file execution. Randomising the time into execution used for training and classification reduces the chances of having a blinkered view of model capabilities.

Without regularisation measures, the representations learned by a neural network can fail to generalise well. For regularisation, we try using dropout as well as \(l1\) and \(l2\) regularisation on the weight and bias terms in the network in our search space. Dropout \cite{srivastava2014dropout} randomly omits a pre-defined percentage of nodes each training epoch, which commonly limits overfitting. \(l1\) regularisation penalises weights growing to large values whilst \(l2\) regularisation allows a limited number of weights to grow to large values. Our random search indicated that a dropout rate of 0.1-0.3 produced the best results on the training set, but weight regularisation was also prevalent in the best-performing configurations. 

Bidirectional RNNs use two layers in every hidden layer, one processing the time series progressively, and the second processing regressively. Pasacnu et al. \cite{PascanuMalwareRNN} found good results using a bidirectional RNN, as the authors were concerned that the start of a file's processes may be forgotten by a progressive sequence as if the LSTM cell forgets it in favour of new data, the regressive sequence ensures that the initial data remains prevalent in decision-making. We also found that many of the the best-scoring configurations used a bidirectional architecture. 

A model depth of 2 or 3 gave the best results. The number of hidden neurons was 50 or more in each layer to give any accuracy above 60\%. All configurations used the ``Adam" weight updating rule \cite{adam} as it learns to adjust the rate at which weights are updated during training. 

\subsection{Predicting malware using early-stage data}\label{early_prediction}

Our goal is to predict malware using behavioural analysis quickly enough that user experience would not (significantly) suffer from the time delay. If the model is accurate within a short time, this sandbox-based analysis could be integrated into an endpoint antivirus system. 

We tested RNNs against other machine learning algorithms used for behavioural malware classification: Random Forest, J48 Decision Tree, Gradient Boosted Decision Trees, Support Vector Machine (SVM), Naive Bayes,  K-Nearest Neighbour and Multi-Layer Perceptron algorithms (as in \cite{tian2010differentiating}, \cite{FirdausiBehaviour}, \cite{wu2014droiddolphin}, \cite{xgboostanew}). Previous research indicates that Random Forest, Decision Tree or SVM are likely to perform the best of those considered.

To mimic the challenge of analysing new incoming samples, we have derived a test set using only the samples that were first seen by VirusTotal after 11:15 on 10th October 2017. This does not account for variants of the same family being present in both the test and training set, but we explore this question in Section~\ref{families}. 

\begin{figure}[h]
\centering
\includegraphics[width=0.95\textwidth]{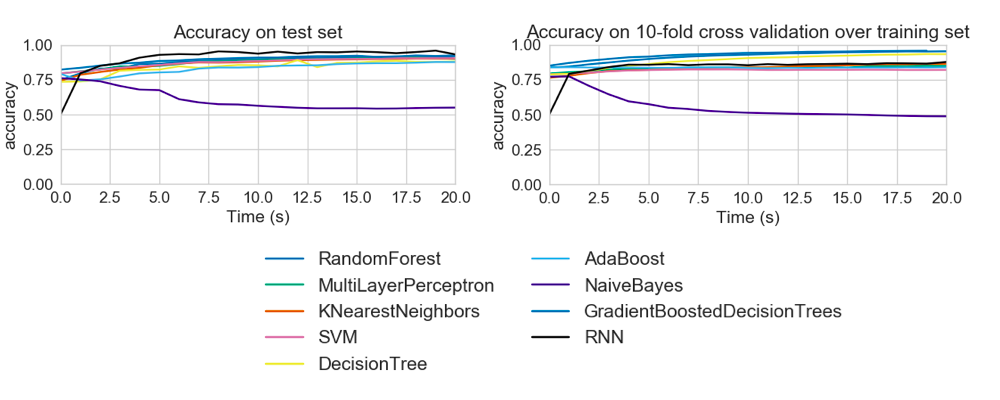}
\caption{Classification accuracy for different machine learning algorithms and a recurrent neural network as time into file execution increases}
\label{other_classifier_graph}
\end{figure}

Figure~\ref{other_classifier_graph} shows the accuracy trend as execution time progresses for the 10-fold cross validation on the training set and on the test set. Random Forest achieves the highest accuracy over the 20 seconds of execution on the training set (see Table~\ref{other_classifiers_10fold}), but the RNN achieves the highest accuracy on the unseen test set (see Table~\ref{other_classifiers_TT}) and outperforms all other algorithms on the unseen test set after 1 second of execution (see lower graph in Figure~\ref{other_classifier_graph}). This could be because the training set is quite homogeneous and so relatively easy for the Random Forest to learn, but it is unable to generalise as well as the RNN to the completely new files in the test set. The RNN cannot usefully learn from 0 seconds as there is no sequence to analyse so accuracy is equivalent to random guess. Using just 1 snapshot (at 0 seconds) of machine activity data, the SVM performs best on the test set and is able to classify 80\% of unseen samples correctly. But after 1 second the RNN performs consistently better than all other algorithms. Using 4 seconds of data the RNN correctly classifies 91\% of unseen samples, and achieves 96\% accuracy at 19 seconds into execution, whereas the highest accuracy at any time predicted by any other algorithm is 92\% (see Table~\ref{rnn_scores}). The RNN improves in accuracy as the amount of sequential data increases. Although peak accuracy occurs at 19 seconds, the predictive accuracy gains per second begin to diminish after 4 seconds . From 0 to 4 seconds accuracy improves by 41 percentage points (11 percentage points from 1 second to 4 seconds) but only by 5 points from 4 to 19 seconds. Our results indicate that dynamic data from just a few seconds of execution can be used to predict whether or not a file is malicious. At 4 seconds we are able to accurately classify 91\% of samples, which constitutes an 8 percentage point loss from the state of the art dynamic detection accuracy\cite{Huang:2016} in exchange for a 04:56 minutes time saved from the typically documented data recording time per sample (see Table~\ref{dynamic_times}), making our model a plausible addition to endpoint anti-virus detection systems.
\begin{table}[h]
    \small
    \centering
    \begin{tabular}{l|l|l|l|l} \hline
        Classifier & Accuracy (\%) & Time (s) & FP (\%) & FN (\%) \\\hline\hline         
        RandomForest & 95.29 & 19 & \textbf{5.03} & 4.5 \\\hline
        MultiLayerPerceptron & 85.01 & 20 & 21.3 & 9.83 \\\hline
        KNearestNeighbors & 86.3 & 20 & 17.53 & 10.96 \\\hline
        SVM & 82.39 & 10 & 24.5 & 10.62 \\\hline
        DecisionTree & 93.41 & 20 & 7.87 & 5.72 \\\hline
        AdaBoost & 83.94 & \textbf{2} & 19.78 & 12.03 \\\hline
        NaiveBayes & 77.44 & \textbf{2} & 29.78 & 10.7 \\\hline
        GradientBoostedDecisionTrees & \textbf{95.81} & 19 & 5.44 & \textbf{3.32} \\\hline
        RNN & 87.75 & 20 & 10.93 & 15.15\\\hline
    \end{tabular}
    \caption{Highest average accuracy over 10-fold cross validation on training set during first 20 seconds of execution with corresponding false positive rate (FP) and false negative rate (FN)}
    \label{other_classifiers_10fold}
\end{table}
\begin{table}[h]
    \small
    \centering
    \begin{tabular}{l|l|l|l|l}
    \hline
        Classifier & Accuracy (\%) & Time (s)  & FP (\%)  & FN (\%)  \\\hline\hline 
        RandomForest & 92.05 & 20 & 4.29 & 12.29 \\\hline
        MultiLayerPerceptron & 91.07 & 18 & 5.53 & 12.98 \\\hline
        KNearestNeighbors & 90.38 & 18 & 4.66 & 15.12 \\\hline
        SVM & 90.57 & 20 & 5.13 & 14.39 \\\hline
        DecisionTree & 89.17 & 12 & 5.22 & 17.22 \\\hline
        AdaBoost & 87.82 & 19 & 7.24 & 17.72 \\\hline
        NaiveBayes & 76.25 & \textbf{0} & 24.74 & 21.13 \\\hline
        GradientBoostedDecisionTrees & 92.62 & 20 & 4.33 & 11.08 \\\hline
        RNN & \textbf{96.01} & 19 & \textbf{3.17} & \textbf{4.72} \\\hline
    \end{tabular}
    \caption{Highest accuracy on unseen test set during first 20 seconds of execution with corresponding false positive rate (FP) and false negative rate (FN)}
    \label{other_classifiers_TT}
\end{table}
\FloatBarrier
\begin{table}[ht]
    \small
    \centering
    \begin{tabular}{p{1cm}|l|l|l|l|l|l|l|l|l|l|l|l|l|l|l|l|l|l|l|l} \hline
    Time (s) & 1&2&3&4&5&6&7&8&9&10&11&12&13&14&15&16&17&18&19&20 \\\hline
    Acc. (\%) & 80&85&87&91&93&94&93&95&95&94&95&94&95&95&95&95&94&95&96&93
    \\\hline
    FN (\%) & 12&14&16&14&10&9&10&5&7&9&6&9&6&7&7&6&9&7&5&7\\\hline
    FP (\%) &33&17&9&2&2&3&2&3&2&2&2&2&4&3&2&4&3&3&3&5
    \\\hline
    \end{tabular}
    \caption{RNN prediction Accuracy (Acc.), false negative rate (FN) and false positive rate (FP) on test set from 1 to 20 seconds into file execution time}
    \label{rnn_scores}
\end{table}

\subsection{Simulation of zero-day malware detection}\label{families}

Dividing the test and training set by date ensures that the two groups are distinct sets of files. However, a slight variant on a known strain is technically a new file. We were able to extract information about the malware families and variants and want to test how well the model performs when confronted with a completely new family or variant. 

Table~\ref{familyOmittable} gives the numbers in the test set for the families and those variants for which there were more than 100 instances in the dataset. Dinwod, Eldorado, Zusy and Wisdomeyes are Trojans; Kazy and Scar are Viruses. We also collected all of those variants listed as advanced persistent threats (APTs) for as signature based systems struggle to detect these especially if previously unseen. The APTs and some of the high-level families have less than 100 samples and as such the results are unlikely to be indicative for the general population of that family but we test them anyway for comparison. 

\begin{table}[h]
    \small
    \centering
    \begin{tabular}{l|r}
    \hline
        Family/Variant     & Total \\\hline\hline
        Trojan             &  1,382 (0)(76)\\
        Virus              &   407 (20)(56)\\
        Adware             &   180 (0)(51)\\
        Backdoor           &   123 (7)(0)\\
        Bot                &   76\\
        Worm               &   24\\
        Rootkit            &   11\\\hline
        Dinwod         &   265 \\
        Artemis        &   228 \\
        Eldorado       &   209 \\
        Zusy           &   135 \\
        Wisdomeyes     &   132 \\
        Kazy           &   116 \\
        Scar           &   101 \\\hline
        APTs            &   27 \\\hline
    \end{tabular}
    \caption{Test accuracy difference between family omitted and included in training set}
    \label{familyOmittable}
\end{table}

To avoid contamination from those samples that were disputed, these are removed from the dataset for the following experiments. For each family in Table~\ref{familyOmittable}, we trained a completely new model without any samples from the family of interest.

The test set is entirely malicious, which means accuracy is an appropriate metric as it is just the rate of correct detection from the only class of interest. Table~\ref{family_results} gives the predictive accuracy over time for different families and for APTs, and Table~\ref{variant_results} gives the predictive accuracies for the five variants for which we collected over 100 samples. Perhaps surprisingly, we see high classification accuracies across these two sets of results. The families are detected with lower accuracy in general. For the Trojans particularly, during the first few seconds, accuracy is actually worse than random chance. Because so much of the dataset set is comprised of Trojans, removing these from training halves the number of malware samples, so this may account for the particularly poor performance. The accuracy does increase significantly between 1 and 3 seconds of execution. This is probably because Trojans are defined by their delivery mechanism, and the model has not been trained on any examples of this form of malware delivery. The model has, however, seen malicious behaviour from other families, which may be similar to some of the later behaviours by the Trojans, accounting for the significant rise in accuracy. Though the Worms are actually detected with a 100\% accuracy at each second, there were only 24 Worm samples in the dataset.

\begin{table}[h]
    \small
    \centering
    \begin{tabular}{p{1.9cm}|l|l|l|l|l|l|l|l|l|l}
    \hline
     Family & \multicolumn{10}{c}{Time(s)} \\\hline
    & 1 & 2 & 3 & 4 & 5 & 6 & 7 & 8 & 9 & 10 \\\hline
Trojan	&		11.16	& 49.67	& 70.23	& 68.07	& 73.86	& 69.33	& 55.63& 57.75	& 60.18	& 56.24	\\\hline
Virus	&		91.26	& 89.58	& 82.7	& 83.0	& 83.54	& 88.89	& 84.56& 86.31	& 84.38	& 82.26	\\\hline
Adware		&90.68	& 90.0	& 83.33	& 84.11	& 59.59	& 85.71	& 87.22& 66.41	& 77.31	& 73.5	\\\hline
Backdoor	&91.3	& 91.21	& 80.0	& 83.53	& 82.28	& 79.73& 87.32	& 82.61	& 79.69	& 80.7	\\\hline
Bot		&93.06	& 91.55	& 92.86	& 84.85	& 90.16	& 85.71	& 80.0	& 86.36	& 88.1	& 87.5	\\\hline
Worm	&		100.0	& 100.0	& 100.0	& 100.0	& 100.0	& 100.0	& 100.0& 100.0	& 100.0	& 100.0	\\\hline
Rootkit	&		100.0	& 75.0	& 75.0	& 75.0	& 100.0	& 75.0	& 100.0& 100.0	& 66.67	& 100.0	\\\hline
APT	& 96.3	& 96.3	& 88.46	& 92.0	& 100.0	& 94.74	& 94.74& 100.0	& 94.74	& 89.47	\\\hline
   \end{tabular}
    \caption{Classification accuracy on different malware families with all instances of that family removed from training set}
    \label{family_results}
\end{table}

\begin{table}[h]
    \small
    \centering
    \begin{tabular}{p{1.9cm}|l|l|l|l|l|l|l|l|l|l}
    \hline
     Variant & \multicolumn{10}{c}{Time(s)} \\\hline
    & 1 & 2 & 3 & 4 & 5 & 6 & 7 & 8 & 9 & 10 \\\hline
    Dinwod	& 90.57	& 89.43 & 78.11	& 91.32	& 93.96	& 98.87	& 99.25	& 98.11 & 98.08	& 97.31	\\\hline
    Eldorado & 94.3 & 93.3	& 92.0	& 86.42	& 90.07	& 82.01	& 74.81	& 81.75 & 85.48	& 83.61	\\\hline
    Wisdomeyes & 92.59 & 90.91	& 83.72	& 91.34	& 89.83	& 92.63	& 94.44	& 84.52 & 90.36	& 87.34	\\\hline
    Zusy	& 91.18	& 89.63	& 85.94	& 82.11	& 81.74	& 85.19	& 85.29	& 88.66 & 90.43	& 85.56	\\\hline
    Kazy	& 89.74	& 82.76	& 85.22	& 86.49	& 87.88	& 94.94	& 87.5	& 88.89 & 91.43 & 89.71	\\\hline
    Scar & 92.08&92.08&75.25&78.22&62.63&81.82&89.69&81.44&86.46&88.42\\\hline

   \end{tabular}
    \caption{Classification accuracy on different malware variants with all instances of that variant removed from training set}
    \label{variant_results}
\end{table}

\begin{figure}[h]
    \centering
    \includegraphics[width=0.5\textwidth]{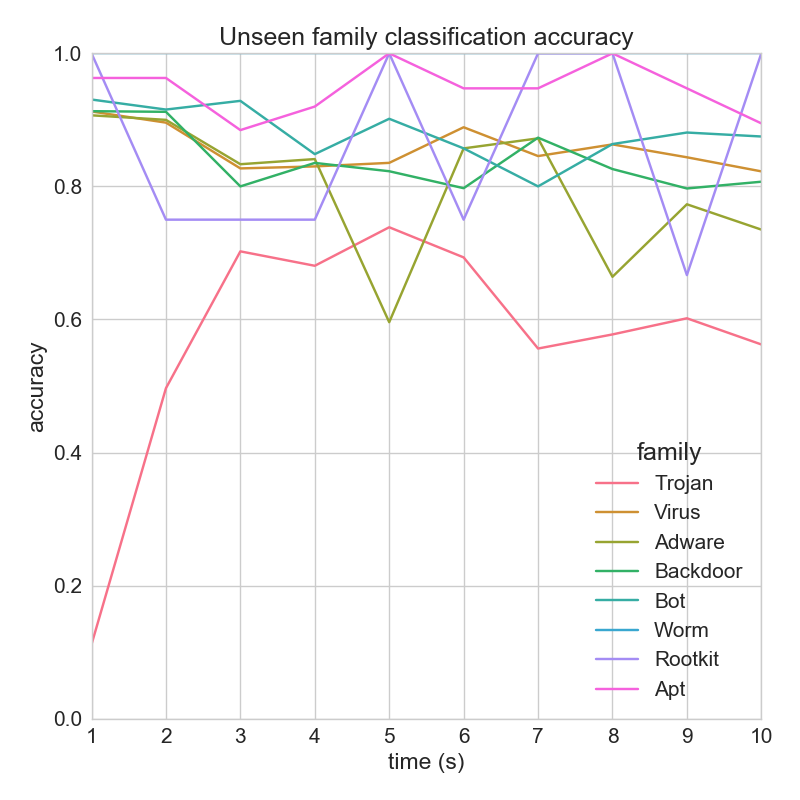}
    \caption{Comparative detection accuracy on various malware families with examples of the family omitted from the training set}
    \label{other_classifier_graph_fam}
\end{figure}

\begin{figure}[h]
    \centering
    \includegraphics[width=0.5\textwidth]{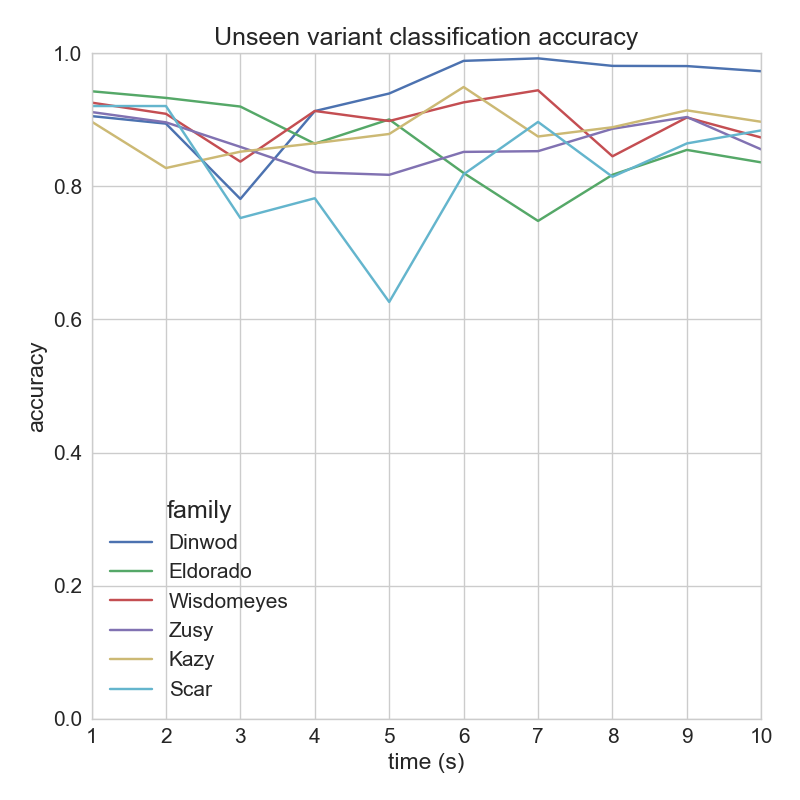}
    \caption{Comparative detection accuracy on various malware variants with examples of the variant omitted from the training sett}
    \label{other_classifier_graph_var}
\end{figure}
\FloatBarrier

The variants tend to achieve a higher predictive accuracy than the families. Other than Dinwod, all families score lower at 10 seconds than at 1 second. Each variant is a kind of Trojan or Virus, but the model was trained on other types of Trojan and Virus. This can help explain the slight drop in accuracy over the first 10 seconds. It is the delivery mechanism which the variants have in common with samples in the training set, so the period over which this occurs (the first few seconds) gives the best predictive accuracy. Every variant was detected with over 89\% accuracy during the first second of execution, despite the model having no exposure to that variant previously. 

If the model is able to score well on a family without ever having seen a sample from that family, the model may hold a robustness against zero days, and support our hypothesis that malware do not exhibit wildly different behavioural activity from one another as their goals are not wildly divergent, even if the attack vector mechanisms are.

\subsection{Ransomware Case Study}\label{ransomware}

Early prediction that a sample is malicious enables defensive techniques to move from recovery to prevention. This is particularly desirable for malware such as ransomware, from which data recovery is only possible by paying a ransom if a backup does not exist. We obtained an additional 2,788 ransomware samples from the VirusShare website \cite{VirusShare} to test the predictive capability of our model. 

Reports in the wake of the high profile ransomware attacks, e.g. WannaCry/WannaDecryptor worm in May 2017, were reported to be preventable if a patch released two months earlier had been installed \cite{naoreport}. Endpoint users cannot be relied on to carry out security updates as the primary defence against new malware. We test our method by removing the 183 ransomware samples and the disputed-family samples from our original dataset and train a new model on the remaining samples, we then test how well the model is able to detect the VirusShare samples and the removed 183 samples. 

The model is able to detect 94\% of samples at 1 second into execution without having seen any ransomware previously. When we include half of the ransomware samples in the training set, this rises to 99.86\% (see Table~\ref{ransomresults}). 

\begin{table}[h]
    \small
    \centering
    \begin{tabular}{p{2.6cm}|l|l|l|l|l|l|l|l|l|l}
    \hline
     Samples in Training Set & \multicolumn{10}{c}{Time(s)} \\\hline
    & 1 & 2 & 3 & 4 & 5 & 6 & 7 & 8 & 9 & 10 \\\hline
    Omitted	& 94.19	& 93.72	& 90.94	& 92.02	& 86.77	& 92.46& 89.55	& 87.62	& 77.88	& 87.52	\\\hline
    Half included & 99.86 & 99.1 & 97.96	& 98.83 & 98.29 & 97.89	& 98.78	& 99.29	& 97.96	& 96.46 \\\hline
    \end{tabular}
    \caption{Classification accuracy on ransomware for one model which has not been trained on ransomware (omitted), and for one which has (half included)}
    \label{ransomresults}
\end{table}

\begin{figure}[h]
\centering
\includegraphics[width=0.75\textwidth]{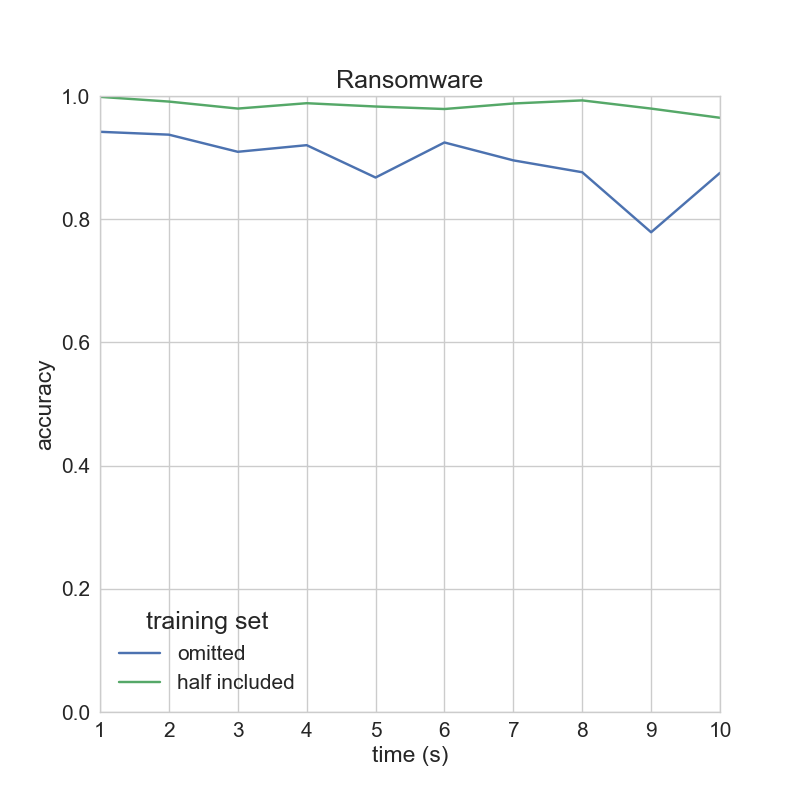}
\caption{Classification accuracy on ransomware for one model which has not been trained on ransomware (omitted), and for one which has (half included)}
\label{ransomware_graph}
\end{figure}

In Figure~\ref{ransomware_graph} there is a clear distinction in the accuracy trend over execution time between the model which has been trained on some of the relevant family. The model which has never seen ransomware before starts to drop in accuracy after the initial few seconds. Again we believe this is because the model is recognising the delivery mechanism at the start of execution, which will be common to other types of malware in the training set, though the later malicious behaviour is is less recognisable to the model by comparison with the later behaviour of the other types of malware it has seen. The model trained with half of the samples knows how ransomware behaves after a few seconds and so maintains a high detection accuracy.

It would be interesting to see if the model at 1 second and the model at 5 seconds rely on different input features to reach accurate predictions. It is difficult to penetrate the decision making process of a neural network; the architecture presented here has 1,344 neurons almost 4 million trainable parameters, but we can turn the input features on and off and see the effect of combinations of features on classification accuracy. By setting the inputs to zero, which is the normalised mean of the training data, we can turn a feature ``off". By turning off all the features and then turning them back on sequentially, we can see which features are needed to gain a certain level of accuracy. 

\begin{table}[h]
    \small
    \begin{tabular}{p{1.3cm}||p{1.3cm}|p{1.5cm}|p{1.3cm}|p{1.5cm}|p{1.3cm}|p{1.5cm}|p{1.3cm}|p{1.5cm}}
    \hline
    & \multicolumn{4}{l|}{Ransomware omitted from training set} & \multicolumn{4}{l}{Ransomware in training set}  \\\hline
    \# Features on & \multicolumn{2}{l|}{1 second model} &\multicolumn{2}{l|}{5 second model} & \multicolumn{2}{l|}{1 second model} &\multicolumn{2}{l}{5 second model} \\\hline

    & Max. Acc. & Features on & Max. Acc. & Features on & Max. Acc. & Features on & Max. Acc. & Features on \\\hline\hline
    1 & 00.03 & tx bytes & 40.82 & memory & 89.36 & rx packets & 14.95 & total processes\\\hline 
    2 & 98.92 & memory and rx bytes & 97.54 & rx bytes and rx packets & 99.80 & tx packets and \{rx packets, rx bytes\} & 71.15 & rx bytes and tx bytes \\\hline
    \end{tabular}
    \caption{Maximum accuracy scores in predicting ransomware with only one and two features turned on for a model not trained on ransomware and for a model trained on ransomware}
    \label{ransomfeaturepower}
\end{table}

In Table~\ref{ransomfeaturepower}, we can see that with just two features, both the 1 second and the 5 seconds models trained with and without ransomware are able to beat 50\% accuracy (random chance). The model trained using ransomware is able to correctly detect more than 99\% of ransomware samples as malicious using just the number of packets sent and either the number of packets or number of bytes received. Unlike the model trained with ransomware, which draws accurate conclusions from packet data and total processes, when no ransomware is included in the training set, memory usage is also a prominent feature in accurate detection. Comparing to the broader families, in classifying Adware, Trojans and Viruses, memory and packets a single input feature allowed the model to achieve more than 50\% accuracy, Trojans are the only family for which memory contributes to scoring above 50\% at the one-second model, when combined with packets sent and swap. As Trojans comprise the majority of the dataset it makes sense that the most relevant features for classifying them help to define what constitutes malware to the model. 

The accuracy in identifying unseen families highlights the presence of shared dynamic characteristics between different malware types. The broad families, which detail the malware infection mechanism particularly help to identify malware early on. Whilst new malware variants are likely to appear, new delivery mechanisms are far less common and help to distinguish unseen families from benignware.  

\subsection{Improving prediction accuracy with an ensemble classifier}\label{ensemble}

As well as accuracy, the values of the model predictions increase with time into file execution. Therefore we now propose an ensemble method, using the top three best performing configurations found in the hyperparameter search space during the previous experiments, to try and improve the classification confidence earlier in the file execution. Accuracy does not increase monotonically in our first configuration, and of the best three configurations on the 10-fold cross-validation, no single configuration consistently achieved the highest accuracy at each second, the configuration used in the previous sections was the configuration that scored the highest accuracy at 1 second. 

We take the best-scoring configurations on the training set across the first 5 seconds, which are 3 distinct hyperparameter sets (one model was the best at 1 and 2 seconds, one at 3 and 5 seconds) and take the maximum of the predictions of these three RNNs before thresholding at 0.5 to give a final malicious/benign label. The configuration details are in Table~\ref{ABCconfig}, configuration ``A" is the same as has been used in the previous experiments. 

\begin{table}[h]
    \centering
    \small
    \begin{tabular}{l|l|l|l}
    \hline
        Hyperparameter          & A & B & C \\\hline\hline
        Depth                   & 3 & 1 & 2 \\\hline
        Bidirectional           & True & True & False \\\hline
        Hidden neurons           & 74 & 358 & 195 \\\hline
        Epochs                   & 53 & 112 & 39 \\\hline
        Dropout rate             & 0.3 & 0.1 & 0.1 \\\hline
        Weight regularisation   & \(l2\) & \(l2\) & \(l1\) \\\hline
        Bias regularisation      & None & None & None \\\hline
        Batch size               & 64 & 64 & 64 \\\hline
    \end{tabular}
    \caption{Highest accuracy-scoring configurations during first 5 seconds in 10-fold cross validation on training set}
    \label{ABCconfig}
\end{table}

To combine the predictions of configurations A, B and C we take the maximum value of the three to bias the predictions in favour of detecting malware (labelled as 1) over benignware (labelled as 0). An ensemble of models does tend to boost accuracy, increasing detection from 92\% to 94\% at 5 seconds, and the maximum accuracy from configuration A alone, 96\%, is reached at 9 seconds instead of at 19 seconds (see Table~\ref{ensembleResults}). The results in Table~\ref{ensembleResults} show that the accuracy score improves or matches the highest scoring model of configurations A, B and C for 12 of the first 20 seconds. Model A, the original configuration, only bests the ensemble accuracy once. We tested whether the ensemble scores improved predictive confidence on the individual samples compared with the predictions of the best-scoring model. We can measure predictive confidence by rewarding those correct predictions closer to 1 or 0 more highly, i.e. a prediction of 0.9 is better than 0.8 when the sample is malicious. The equation for predictive confidence is as follows: \[confidence = 1 - |b - p|\] where \(b\) is the true label and \(p\) is the predicted label.

\begin{table}[ht]
    \small
    \centering
    \begin{tabular}{l|p{3cm}|l|l|l}
    \hline
    Time (s) & Highest accuracy of configurations A, B and C & Ensemble acc. (\%) & Ensemble FP (\%) & Ensemble FN (\%) \\\hline
    1 & \textbf{79.69} (C) & 79.5 & 33.5 & 12.03 \\\hline
    2 &\textbf{85.6} (A) & 83.69* & 25.73 & 10.16 \\\hline
    3 & 87.52 (A, C) & \textbf{88.48*} & 15.05 & 9.21 \\\hline
    4 & 91.54 (A) & \textbf{91.92*} & 8.74 & 7.64 \\\hline
    5 & 92.38 (B) & \textbf{93.95*} & 3.4 & 7.84 \\\hline
    6 & 94.09 (A) & \textbf{95.28*} & 4.37 & 4.97 \\\hline
    7 & 94.92 (A) & \textbf{95.12*} & 4.85 & 4.9 \\\hline
    8 & 94.25 (A) & \textbf{95.48*} & 4.88 & 4.26 \\\hline
    9 & 94.97 (A) & \textbf{96.02*} & 4.39 & 3.68 \\\hline
    10 & \textbf{95.53} (C) & 95.11* & 5.45 & 4.48 \\\hline
    11 & 95.91 (C) & \textbf{96.13*} & 4.95 & 3.04 \\\hline
    12 & \textbf{95.46} (C) & \textbf{95.46*} & 5.47 & 3.82 \\\hline
    13 & 95.16 (A) & \textbf{95.6*} & 5.97 & 3.15 \\\hline
    14 & \textbf{95.93} (C) & \textbf{95.93*} & 5.03 & 3.29 \\\hline
    15 & \textbf{96.1} (C) & 95.87* & 4.57 & 3.77 \\\hline
    16 & 95.62 (C) & \textbf{96.54*} & 4.08 & 2.94 \\\hline
    17 & 95.34 (C) &\textbf{96.5*} & 3.06 & 3.86 \\\hline
    18 & \textbf{96.67} (C) & 96.43* & 4.12 & 3.1 \\\hline
    19 & \textbf{96.51} (C) & 96.26* & 4.23 & 3.3 \\\hline
    20 & 93.81 (A) & \textbf{94.85}* & 8.22 & 3.31 \\\hline
    \end{tabular}
    \caption{Ensemble accuracy (acc.), false positive rate (FP) and false negative rate (FN) compared with highest accuracy of configurations A, B and C. Those marked with a ``*" signify predictions that were statistically significantly more confident by at the confidence level of 0.01 }
    \label{ensembleResults}
\end{table}

Using a one-sided T-test, we found that the confidence of predictions from the ensemble method were significantly higher (at 0.01 confidence level) for every second after 1 second, malicious predictions are likely be more confident as we are taking the maximum value of the three models, but it is interesting that taking the maximum of the benign samples does not out weigh the increase in confidence. This indicates that three models are more confident about benign samples than malicious ones. A further benefit of the ensemble approach is the reduction in the false negative rate. The minimum false negative rate for Model A was 4.5\%, but here the false positive rate is at least 3 percentage points lower than for model A during the first 7 seconds, and remains lower than Model A's global minimum for the remaining 20 seconds. 

If the gains in accuracy for the ensemble classifier are due to differences in the features learned by the network, this could help to protect against adversarial manipulation of data. We attempt to interpret what configurations A, B, and C are using to distinguish malware and benignware. These preliminary tests seek to gauge whether it is possible to analyse the decisions made by the trained neural networks. 

By setting the test data for a feature (or set of features) to zero, we can approximate the absence of that information between samples. We  assess  the overall impact of turning features ``off” by observing the fall in accuracy and dividing it by the number of features turned off. A single feature incurring a 5 percentage point loss attains an impact factor of -5, but two features creating the same loss would be awarded -2.5 each. Finally, we take the average across impact scores to assess the importance of each feature when a given number of features are switched off.

\begin{figure}[h]
      \centering
        \begin{tabular}{@{}c@{}}
      \includegraphics[width=0.6\linewidth]{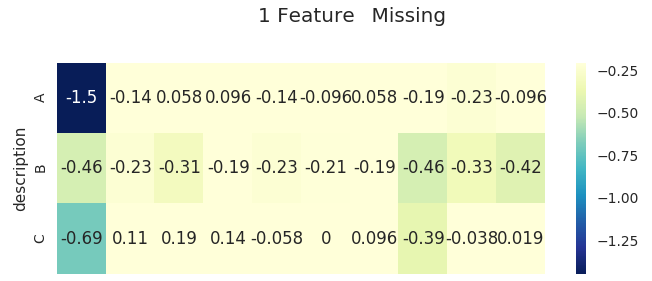}
        \end{tabular}
    
      \vspace{\floatsep}
    
      \begin{tabular}{@{}c@{}}
      \includegraphics[width=0.6\linewidth]{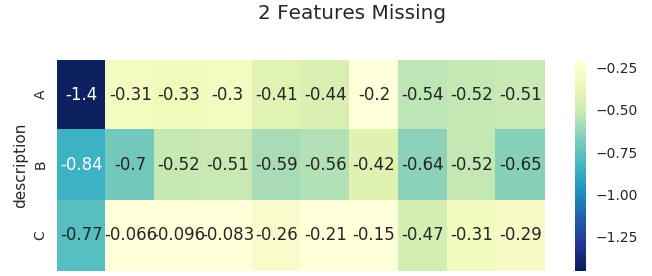}
        \end{tabular}
        
        \begin{tabular}{@{}c@{}}
      \includegraphics[width=0.6\linewidth]{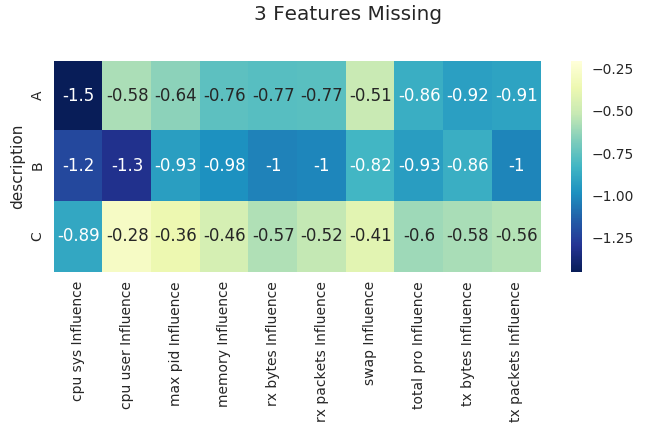}
        \end{tabular}
        
    \caption{Impact scores for features with 1, 2 and 3 features turned off 4 seconds into file execution}
    \label{leave_out_graph}
\end{figure}

Figure~\ref{leave_out_graph} gives the impact factors for each feature at 4 seconds into file execution. Intuitively, the more features omitted, the higher the impact factors become. Interestingly, there are some very small gains in accuracy for configurations A and B when only one feature is missing but no more than 0.2 percentage points. For each of the configurations, CPU use on the system has the highest impact factor. It is most integral for configuration A, which is also the best-scoring model. The CPU use in configuration A does not really see an increase in its impact factor as we remove more input features, but for configuration B, all features attain higher impact factors the more are removed. We can infer that configuration B has learned a representation of the data which combines the inputs to decide whether the output is malicious or benign, whereas configuration A appears to have learned at least one representation of CPU system use as a predictor of malware.  

The difference between the impact scores and their emphasis can help us to see which features are most predictive at different time steps (at 4 seconds this is CPU usage) and to understand how an ensemble classifier is able to outperform the predictions of its components. As all three models suffer the biggest loss from CPU usage, if an adversary knew this she might be able to manipulate CPU system use to avoid detection. Future work should examine the decision processes of networks to detect potential weaknesses that could be exploited to evade detection. The ensemble offers a small increase in accuracy but more importantly, this analysis can help to understand ways in which the models may be manipulated, by biasing results towards malicious predictions (taking the maximum prediction) we introduce a form of safety-net against the manipulation of a single model. 

\section{Limitations and Future work}

Our results indicate that behavioural data can provide a good indication of whether or not a file is malicious based only on its initial behaviours, even when the model has not been exposed to a particular malware variant before. Dynamic analysis could reasonably be incorporated into endpoint antivirus systems if the analysis only takes a few seconds per file. Further challenges which must be addressed before this is possible include:

\subsection{Other file types and operating systems}

So far we have only examined Windows7 executables. Though Windows7 is the most prevalent operating system globally \cite{win7} and Windows executables are the most commonly submitted file to VirusTotal \cite{virusTotalStats}, we should extend these methods to see if the model is capable of detecting malicious PDFs, URLs and other potential vehicles for malware, as well as applications which run on other operating systems.

\subsection{Robustness to adversarial samples}

The robustness of this approach is limited if adversaries know that the first 5 seconds are being used to determine whether a file will run in the network. By planting long sleeps or benign behaviour at the start of a malicious file, adversaries could avoid detection in the virtual machine. We hypothesised that malicious executables begin attempting their objectives as soon as possible to mitigate the chances of being interrupted, but this would be likely to change if malware authors knew that only subsections of activity were the basis of anti-virus system decisions. We envisage future work examining a sliding-window approach to behavioral prediction.

The sliding-window approach will take snapshots (of 5 seconds) of data and monitor machine activity on a per-process basis to try and predict whether or not a file is malicious. This would run in the background as the file is executed in a live environment. The advantage of this approach is that we eliminate the waiting time before a user is allowed to access the file. The challenges in implementing these next steps are recalibration for endpoint machines (see Section \ref{machineactdiscuss} below) and sufficiently quick killing of the malicious process once it has been detected, i.e. before the malicious payload is executed.

Despite the future worry that executables could be amended to avoid detection by the model proposed in this paper, this does not invalidate the use of our proposed method. Whilst some attacks may be altered specifically to evade an behavioral early-detection system, this would be in response the attacker knowing that the target in question was employing these types of defence. However, there would still be many malwares without benign behaviour injections at the start of the file. We continue to use signature-based detection in antivirus systems despite the use of static obfuscation techniques, because it is still an invaluable method for quickly detecting previously seen malwares. The model proposed here indicates that we can quickly detect unseen variants, and we hope that future research will evaluate the robustness of the sliding window approach using adversarially crafted samples.

\subsection{Process blocking}

If a live monitoring system is implemented, processes predicted to be malicious will need to be terminated. Future work should examine the ability of the model to block once the classifier anticipates malicious activity, and to investigate whether the malicious payload has been executed.  

\subsection{Portability to other machines and operating systems}\label{machineactdiscuss}

The machine activity metrics are specific to the context of the virtual machine used in this experiment. To move towards adoption in an endpoint anti-virus system, the RNN should be retrained on the input data generated by a set of samples on the target machine. Though this recalibration will take a few hours at the start of the security system installation, it will only need to be performed when hardware is upgraded (once per machine for most users) and opens the possibility of porting the model to other operating systems, including other versions of Windows.

Though we have not tested the portability of the data between machines, i.e. training with data recorded on one machine and testing with data recorded on another, it is easy to see cases in which this will not work. Some metrics, such as CPU usage are relative (measured as a percentage of total available processing power) and so will change dramatically with hardware capacities. For example, a file requiring 100\% of CPU capacity on one machine may use just 30\% on another with more cores. However, we see no reason why the model cannot be re-calibrated to a new machine. There is cause for concern if the hardware means that the granularity of the data falls below that which is used in this paper. For example a very small amount of RAM could limit the memory usage such that the useful information that one sample uses 1.1MB and another 1.2MB are both capped at 1MB, thus appearing the same to the model. Whilst the experiments in this paper are conducted in a virtual machine and the memory, storage and processing power can be replicated, we hope that future work will extend this model to run live in the background on the intended recipient machine. Since the hardware capacities of a typical modern computer are greater than those for the virtual machine used here, this may in turn provide more granularity in the data and possibly allow the model to learn a better representation of the difference between malicious and benign software. The different results that we would be likely to see on a more powerful machine offer a potential advantage in training but also necessitate re-calibration on a per-machine basis. Since this is a one-off time cost, it is not a major limitation of the proposed solution.

\section{Conclusions}

Dynamic malware detection methods are often preferred to static detection  as the latter are particularly susceptible to obfuscation and evasion when attackers manipulate the code of an executable file. However, dynamic methods previously incurred a time penalty due to the need to execute the file and collect its activity footprint before making a decision on its malicious status. This meant the malicious payload had likely already been executed before the attack was detected. We have developed a novel malware prediction model based on recurrent neural networks (RNNs) that significantly reduces dynamic detection time, to less than 5 seconds per file, whilst retaining the advantages of a dynamic model. This offers the new ability to develop methods that can predict and block malicious files before they execute their payload completely, preventing attacks rather than having to remedy them.

Through our experimental results we have shown that it is possible to achieve a detection accuracy of 94\% with just 5 seconds of dynamic data using an ensemble of RNNs and an accuracy of 96\% in less than 10 seconds, whilst typical file execution time for dynamic analysis is around 5 minutes.

The best RNN network configurations discovered through random search each employed bidirectional hidden layers, indicating that making use of the input features progressing as well as regressing in time aided distinction between malicious and benign behavioural data. 

A single RNN was capable of detecting completely unseen malware variants with over 89\% accuracy for the 6 different variants tested at just 1 second into file execution. The accuracy tended to fall a little after the first 2 seconds, implying that the model was best able to recognise the infection mechanism at a family level (e.g. Trojan, Virus) given that this would be the first activity to occur. The RNN was less accurate at detecting malware at a family level when that family had been omitted from the training data (11\% accuracy at 1 second detecting Trojans), further indicating that the model was easily able to detect new variants, provided it had been exposed to examples of that family of infection mechanisms. Our ransomware use case experiment supported this theory further, as the RNN was able to detect ransomware, which shares common infection mechanisms with other types of attack such as Trojans, with 94\% accuracy, without being exposed to any ransomware previously. However, this accuracy fell as time into file execution increased, again implying that the model was easily able to detect a malicious delivery mechanism, better than the activity itself. After exposure to ransomware, the model accuracy remained above 96\% for the first 10 seconds.

The RNN models outperformed other machine learning classifiers in analysing the unseen test set, though the other algorithms performed competitively on the training set. This indicates that the RNN was more robust against overfitting to the training set than the other algorithms and had learnt a more generalisable representation of the difference between malicious and benign files. This is particularly important in malware detection as adversaries are constantly developing new malware strains and variants in an attempt to evade automatic detection.

To date this is the first analysis of the extent to which general malware executable files can be predicted to be malicious during its execution rather than using the complete log file post-execution, we anticipate that future work can build on these results to integrate file-specific behavioural detection into endpoint anti-virus systems across different operating systems.

\bibliography{main}
\bibliographystyle{elsarticle-num}

\end{document}